# Coulomb explosion of "hot spot".

Oreshkin V.I., Oreshkin E.V, Chaikovsky S.A., Artyomov A.P.

The study presented in this paper has shown that the generation of hard x rays and high-energy ions, which are detected in pinch implosion experiments, may be associated with the Coulomb explosion of the hot spot that is formed due to the outflow of the material from the pinch cross point. During the process of material outflow, the temperature of the hot spot plasma increases, and conditions arise for the plasma electrons to become continuously accelerated. The runaway of electrons from the hot spot region results in the buildup of positive space charge in this region followed by a Coulomb explosion. The conditions for the hot spot plasma electrons to become continuously accelerated have been revealed and estimates have been obtained for the kinetic energy of the ions generated by the Coulomb explosion.

A $Z$ pinch is an electrical discharge in plasma, which is compressed under the action of the magnetic pressure produced by the intrinsic discharge current [1-7]. Typical of Z pinches is the formation of hot spots due to the development of large-scale MHD instabilities [1,2,8-10]. The discharge plasma column is deformed during compression, which is accompanied by the formation of necks smaller in radius than the main column. The magnetic pressure in the neck region increases, resulting, first, in faster compression and, second, in material outflow from the neck region in the axial direction. The final stage of the necking is a hot spot.

Apparently, the axial jets were first detected in an experimental study [11] where the process of compression of deuterium pinches was investigated. As these jets were accounted for by an increase in neutron yield, the authors of Ref. [11] proposed a "noncylindrical Z pinch" in which the formation of the hot spot was predetermined by the geometry of the discharge. Subsequently, this configuration was called a plasma focus [12,13]. In subsequent experiments on a deuterium plasma focus at currents of about 1 MA, electron beams, hard x rays, and "epithermal" deuterons with energies up to 8 MeV were detected [14]. Even more energetic deuterons with energies of several tens of megaelectron-volts were detected in experiments with deuterium liners imploded at a current of 2.7 MA [15]. To explain the generation of high-energy ions and hard x rays, the so-called "target" mechanism was proposed [16]. It is assumed that in the final stage of formation of the hot spot, displacement currents occur that generate strong electric fields in which charged particles are accelerated.

Although hot spots are frequently observed in all varieties of Z pinch, such as plasma focuses [12-14,17], plasma liners [10,14], and vacuum sparks [18], the process of their formation in an X pinch has received the most study. An X pinch is a device consisting of two or more crossed wires (forming the letter "X"), and, like the plasma focus, it was proposed [19] to produce hot spots. In contrast to a plasma focus, where the place of appearance of a hot spot is not known in advance, in an X pinch it is predetermined well enough: this is the wire cross point. This feature of X pinches served as the key to the success of experimental studies of the evolution of the neck and the hot spot [20,21].

In an X pinch, the hot spot is formed according to the following scenario [20-22]. Within a short time after the start of the current flow through the pinch, an electric explosion of the wires occurs. Thereafter, a neck is formed whose transverse dimension is several hundreds of micrometers, and then a hot spot several micrometers in size is formed on the neck. This hot spot serves as a source of soft x rays. At the end, the hot spot explodes and at this time, hard x rays and high-energy ions are generated. The ions appear with a slight delay after the main pulse of soft x rays [21].

This scenario is illustrated by Figs. 1 and 2 that present photographic images of an X pinch at different times (Fig. 1) and the waveforms of the current flowing through the pinch and of the soft-x-ray pulse (Fig. 2). The experiment was carried out on a compact high-current generator [23], which allowed switching a current of amplitude up to 230 kA and rise time 180 ns in the short circuit mode. The diagnostic equipment of the generator consisted of Rogowski coils used to measure current waveforms; vacuum x-ray diodes that allowed recording the pulse waveform of soft x rays with photon energies $h\nu > 1$ keV, and an HSFC Pro optical camera capable of producing four frames in visible light for each shot. The generator was loaded by X pinches, each consisting of two molybdenum wires of diameter 12.7 mm inclined at an angle $\varphi \approx 32°$ to the X-pinch axis. The exposure time of each of the images shown in Fig. 1 was 3 ns.

In Fig. 1*a*, which presents an image of the X pinch taken at peak radiation intensity in the soft-x-ray range (Fig. 2), we can clearly see a neck of length 250 µm and a significantly smaller hot spot. In the second image (see Fig. 1*a*), made 20 ns later, a cavity is seen at the neck place that was formed obviously due to the explosion of the hot spot. Similar pictures were obtained [20,21] for X pinches also consisting of two molybdenum wires of diameter 17 µm the current through which at the time of the explosion was 150–170 kA. These images, taken in soft x rays, clearly demonstrate the explosion of the hot spot. The cause of the explosion can be fast runaway of electrons from the hot spot region, resulting in the buildup of positive space charge followed by a Coulomb explosion. This scenario is supported by the presence of an electron beam, indicating the presence of runaway electrons in the pinch plasma [24,25]. It should be noted that the runaway electrons in gas discharges [26,27] are probably generated also due to a Coulomb explosion. However, in this case, the Coulomb explosion occurs as a consequence of the formation of a negative charge at the head of the anode-directed streamer [28,29].

Next, we show that conditions for the runaway of electrons arise during the evolution of the hot spot. The formation of the neck in a Z pinch was studied in detail [8], and a simple model of this process was proposed. According to this model, the neck is formed during compression of the pinch wires, which is accompanied by the material outflow from the neck region along the axis (in both the positive and the negative direction). This takes place in conditions close to the Bennett equilibrium [30], that is, when the magnetic pressure of the plasma is approximately equal to its thermal pressure. It was shown [31] that for any current density distribution inside a pinch, the Bennett equilibrium condition is given by

$$kT = \frac{I^2}{2c^2 N_i (1+\bar{z})}, \qquad (1)$$

where $I$ is the current through the pinch, $N_i$ is the number of ions per unit length, $\bar{z}$ is the mean ion charge, $T$ is the pinch temperature, $k$ is Boltzmann's constant, and $c$ is the velocity of light in vacuum. In an X pinch, the Bennett equilibrium condition (1) seams to hold up to high degrees of compression. For an aluminum X pinch, the parameters of the hot spot were determined for the time of formation of the soft-x-ray pulse, and it was shown that at this time, condition (1) was fulfilled with good precision [32].

As the material flows out from the neck region, the plasma temperature increases, in accordance with (1), and conditions may arise for the onset of continuous acceleration of the plasma electrons, such that an electron, being in an electrostatic field, gains more energy within the mean free path than loses it in inelastic collisions. The critical electrostatic field strength, $E_{cr}$, at which the friction force fails to balance the electric force at any value of the directed electron velocity is determined by the Dreicer criterion [24]

$$E_{cr} \approx 0.2 \cdot \frac{e\Lambda}{D^2}, \qquad (2)$$

where $e$ is the electron charge, $\Lambda$ is the Coulomb logarithm, and $D$ is the Debye screening radius, which, for multicharge plasma, can be written as [33]

$$D = \sqrt{\frac{kT}{4\pi e^2 n_i \left(\bar{z} + \langle z^2 \rangle\right)}} \quad , \tag{3}$$

where $n_i$ is the ion density in the pinch plasma and $\langle z^2 \rangle \approx \bar{z}^2$ is the RMS ion charge. Below, we assume that $n_i \approx \frac{N_i}{\pi R_{hs}^2}$, where $R_{hs}$ is the radius of the hot spot.

The runaway electron beam is formed in a pinch when the electric field strength in the hot spot, $E_{hs}$, produced by the current flowing through the pinch becomes higher than $E_{cr}$, i.e.

$$E_{cr} < E_{hs} = \frac{I}{\pi R_{hs}^2 \sigma} , \tag{4}$$

where $\sigma$ is the conductivity of the hot spot plasma. The well-known and often used expression for the conductivity of completely ionized plasma across the magnetic field was obtained by Braginsky [34]; it reads

$$\sigma_{Brag} = \frac{3}{4e^2 \sqrt{2\pi m_e}} \frac{(kT)^{3/2}}{\Lambda \bar{z}} \quad , \tag{5}$$

where $m_e$ is the electron mass. However, formula (5) takes into account only elastic electron-ion collisions, and, as noted by Braginsky [34], the actual conductivity of the plasma should be lower due to plasma waves, microturbulence, inelastic collisions, etc. The causes for anomalous resistivity in pinches were discussed, for instance, in Ref. 8; however, there is no commonly opinion on this problem. It was supposed [8] that the decrease in plasma conductivity due to anomalous effects becomes significant when the current electron velocity $u_e = \frac{I}{e\bar{z}N_i}$ becomes greater than the thermal velocity of the ions, $u_i = \sqrt{\frac{2kT}{m_i}}$, where $m_i$ is the atomic mass of the pinch material, i.e. when $u_i < u_e$. In view of (1), the last inequality can be reduced to the condition for the number of ions per unit length:

$$N_i^{an} < \frac{m_i c^2 (1 + \bar{z})}{e^2 \bar{z}^2} . \tag{6}$$

The lower threshold of $N_i^{an}$ below which the anomalous effects become significant is of the order of $10^{16}$ cm$^{-1}$. The temperature at which the inequality $u_i < u_e$ holds is given by

$$kT^{an} = \frac{1}{2} m_e c^2 \left(\frac{m_e}{m_i}\right) \left(\frac{\bar{z}}{1+\bar{z}} \frac{I}{I_A}\right)^2, \tag{7}$$

where $I_A = \frac{m_e c^3}{e} \approx 17$ kA is the Alfven current. In the experiment with a molybdenum pinch the results of which are presented in Figs. 1 and 2, the temperature $kT^{an}$ was about 30 eV.

The impact of anomalous effects can be taken into account [8,35] by presenting the plasma conductivity as $\sigma = \sigma_{Brag}/a$, where $a$ is a dimensionless factor; its value depends on the type of microturbulence structures that scatter electrons [35]. Early in the process, when $N_i > N_i^{an}$, the factor $a$ is close to unity. However, as more and more material leaves the neck region, so that the number of ions in the cross-section of the plasma column decreases to $N_i < N_i^{an}$, the factor $a$ increases. Once this takes place, the plasma starts rapidly heating up, its radial compression slows down [8,17], and the velocity of outflow along the z-axis increases. Approximately at this point in time, the hot spot is formed. As a result of the increase in temperature, the Debye screening radius $D$ increases, the critical electric field strength $E_{cr}$ decreases, and inequality (4) is fulfilled; that is, we have $E_{cr} < E_{hs}$. Using the Bennett condition (1) for the temperature and taking into account relations (2), (3), and (5), we can rewrite inequality (4) as

$$N_i^{cr} < 3.5 \frac{m_e c^2}{e^2} \frac{a^2}{1+\bar{z}}, \tag{8}$$

where $N_i^{cr}$ is the threshold number of ions per unit length of the pinch at which conditions arise for electrons to become continuously accelerated (for $\bar{z} = 20$ and $a = 10$, we have $N_i^{cr} \approx 5 \cdot 10^{13}$ cm$^{-1}$). Comparing inequalities (6) and (8), we find that $N_i^{an}/N_i^{cr} \approx 0.3 \frac{m_i}{m_e} a^2$, whence we obtain the range for the dimensionless factor $a$: $1 < a << 0.5\sqrt{\frac{m_i}{m_e}} \approx 20\sqrt{A}$, where $A$ is the atomic weight of the pinch material measured in atomic mass units.

The runaway electron beam formed in the hot spot region if condition (8) is fulfilled is responsible for the buildup of positive space charge in the pinch plasma. The Coulomb explosion probably occurs approximately at the time when the electrostatic energy of the ions whose electric charge is not balanced by the charge of the electrons becomes greater than the magnetic energy, i.e. $\mathrm{E}^{elec} = \int \frac{E^2}{8\pi} dV \geq \mathrm{E}^{mag} = \int \frac{H^2}{8\pi} dV$, where $E$ and $H$ are, respectively, the electric and the magnetic field strength at the hot spot (the integral is evaluated over the volume in which the positive space charge is built up). The electrostatic energy of the ions in the Coulomb explosion can be estimated using Maxwell's equation $div\mathbf{E} = 4\pi e \bar{z} n_i^+$, where $n_i^+$ is the density of the ions whose electric charge is not balanced by the charge of the electrons. For the case where the radial and the axial dimension of the hot spot are approximately equal to each other, i.e. $R_{hs} \approx L_{hs}$, the following estimate is valid:

$$\mathrm{E}^{elec} \approx \frac{1}{4}\left(e\bar{z}N_i^+\right)^2 L_{hs}. \tag{9}$$

where $N_i^+$ is the number of unbalanced ions per unit length. For a uniform current distribution, the magnetic energy concentrated in the space occupied by the hot spot is given by

$$\mathrm{E}^{mag} = \frac{1}{4}\frac{I^2}{c^2}L_{hs}. \tag{10}$$

Hence, the number of unbalanced ions per unit length should be

$$N_i^+ \approx \frac{m_e c^2}{e^2} \frac{I}{\bar{z} I_A}. \tag{11}$$

As the explosion occurs, the electrostatic energy is converted into the kinetic energy of the hot spot ions, and the average kinetic energy per ion is given by

$$\mathrm{E}_i^{kin} \approx \frac{\mathrm{E}^{ex}}{N_i^{cr} L_{hs}} \approx 0.07 m_e c^2 (1+\bar{z})\left(\frac{I}{aI_A}\right)^2. \tag{12}$$

In the experiment with a molybdenum X pinch the results of which are presented in Figs. 1 and 2, the average kinetic energy of ions in the Coulomb explosion should be about 200 keV (for $a = 10$ and $\bar{z} = 20$). In the experiment [15] with deuterium Z pinches imploded at a current of 2.7 MA, in which, as mentioned, the energy spectrum of fast ions was measured, deutrons with energies of up to 40–50 MeV were detected. Supposing that the average kinetic energy of deutrons $\mathrm{E}_i^{kin}$ was about 20 MeV, we obtain that, according to (12), the dimensionless factor $a$ for this experiment was about 10.

Thus, the hard x rays, the electron beam, and the high energy ions observed in pinch implosion experiments may be related to the Coulomb explosion of the hot spot. The hot spot is formed due to the outflow of the pinch material in the axial direction. As more and more material leaves the hot spot region, the temperature of the hot spot plasma increases, and there arise conditions for the plasma electrons to become continuously accelerated. The electron runaway from the hot spot region results in the buildup of positive space charge that is followed by a Coulomb explosion.

The work was supported by the Russian Science Foundation, grant number 16-19-10142, work Oreshkin E.V. supported by grant RFBR № 16-38-60199.

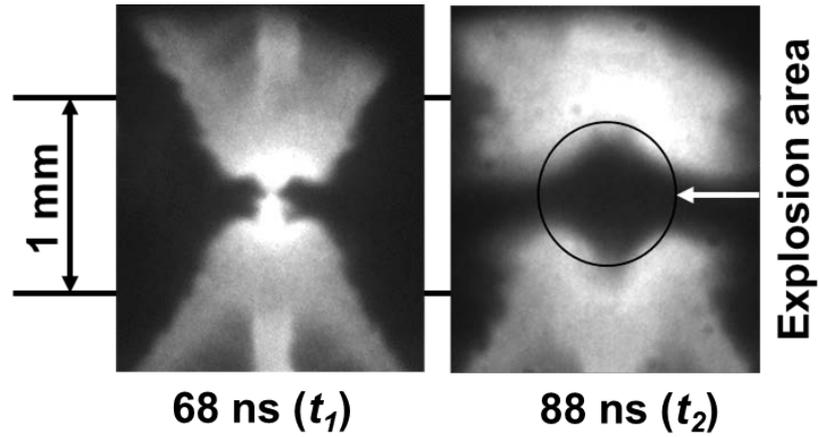

Fig. 1 Images of an X pinch (two molybdenum wires of diameter 12.7 µm) taken with an HSFC-Pro four-frame optical camera at $t_1 = 68$ ns (*a*) and $t_2 = 88$ ns (*b*); an x-ray pulse was detected at $t = 67 \pm 2$ ns.

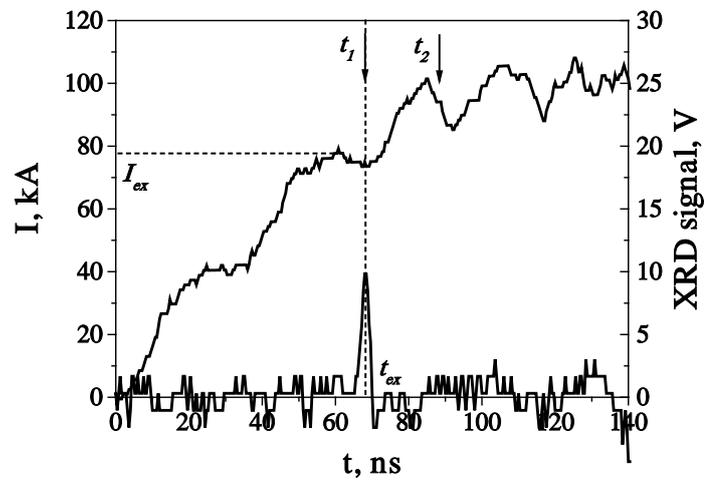

Fig. 2 Waveforms of the current through the X pinch and of the x-ray pulse generated by the pinch.